\documentclass{ptephy_v1}

\preprintnumber{XXXX-XXXX} 

\usepackage{arydshln}
\usepackage{color}
\usepackage{multirow}

\begin{document}

\title{Improved method for measuring low concentration radium and its application to the Super-Kamiokande Gadolinium project}


\author[1,*]{S. Ito}
\affil{Okayama University, Faculty of Science, Okayama 700-8530, Japan \email{s-ito@okayama-u.ac.jp}}

\author[2,3,4,*]{K. Ichimura}
\affil{Kamioka Observatory, Institute for Cosmic Ray Research, University of Tokyo, Kamioka, Gifu 506-1205, Japan \email{ichimura@awa.tohoku.ac.jp}}
\affil[3]{Kavli Institute for the Physics and Mathematics of the Universe (WPI), the University of Tokyo, Kashiwa, Chiba, 277-8582, Japan}
\affil[4]{Present address: Research Center for Neutrino Science, Tohoku University, Sendai 980-8578, Japan}

\author[5]{Y. Takaku}
\affil[5]{Institute for Environmental Sciences, Department of Radioecology, Aomori, 039-3212, Japan}

\author[2,3]{K. Abe}

\author[1]{M.~Harada}

\author[2,3]{M.~Ikeda}

\author[2]{H.~Ito}

\author[2,3,4]{Y.~Kishimoto}

\author[2,3]{Y.~Nakajima}

\author[2]{T.~Okada}

\author[2,3]{H.~Sekiya}


\begin{abstract}%

Chemical extraction using a molecular recognition resin named ``Empore Radium Rad Disk" was developed to improve sensitivity for the low concentration of radium (Ra). 
Compared with the previous method, the extraction process speed was improved by a factor of three and the recovery rate for $^{226}$Ra was also improved from 81$\pm$4\% to $>$99.9\%. 
The sensitivity on the 10$^{-1}$~mBq level was achieved using a high purity germanium detector. 
This improved method was applied to determine $^{226}$Ra in Gd$_2$(SO$_4$)$_3{\cdot}$8H$_2$O which will be used in the Super-Kamiokande Gadolinium project. 
The improvement and measurement results are reported in this paper. 

\end{abstract}

\subjectindex{XXX}

\maketitle

\section{Introduction}

Detection of radium (Ra) is important not only for environmental studies, medical science, biology, but also for non-accelerator particle physics experiments (e.g. neutrino measurements and dark matter searches), which require a low background experimental environment. 
In general, short lifetime radioactive materials (e.g. $^{226}$Ra) are measured using particle counters such as $\alpha$-spectrometers, liquid scintillators, and germanium detectors. 
However, the sensitivity is generally limited by the detection efficiency related to the sample size, i.e. self-shielding effect and geometrical acceptance. 
To minimize these problems and improve sensitivity, chemical extraction is often used. 
For example, Dulansk$\acute{\rm a}$ et {\it al}., \cite{Sr} have used a molecular recognition resin named ``AnaLig-Sr01", which is a product of IBC Advanced Technologies \cite{IBC}. 
They determined the concentration of $^{226}$Ra included in rocks or building materials in the range of $5.2-165.0$ Bq kg$^{-1}$. 
Other chemical extraction techniques have been developed using ``Empore Radium Rad Disk" produced by 3M Corporation \cite{3M} (see the following sections in more details) to determine $^{226}$Ra in water at the  $10^{2}-10^{4}$~mBq~L$^{-1}$ level \cite{Disk1, Disk2, Disk3}. 
However, the chemical extraction of $^{226}$Ra from a very high matrix sample with a much lower concentration has not been established.

The Super-Kamiokande Gadolinium (SK-Gd) project is an upgrade of the Super-Kamiokande (SK) detector  \cite{NIMA}, with its final goal of dissolving gadolinium sulfate octahydrate (Gd$_2$(SO$_4$)$_3 \cdot$8H$_2$O) into the SK detector up to the 0.2\% concentration \cite{Gd, EGADS}. 
One of the main physics targets of SK-Gd is to discover supernova relic neutrinos and study star formation of the universe \cite{SRN}. 
The measurements of solar neutrinos with a low energy threshold of $\sim$3.5~MeV \cite{solar} will be also continued in SK-Gd: therefore several radio impurities (e.g. $^{226}$Ra, $^{238}$U, and $^{232}$Th) in Gd$_2$(SO$_4$)$_3{\cdot}$8H$_2$O should be minimized before loading into SK. 
The maximum allowed level of these radio impurities in Gd$_2$(SO$_4$)$_3{\cdot}$8H$_2$O \cite{Pablo, icrc} and the typical example of a commercially available product are summarized in Table \ref{tab:level}. 
For the measurement of $^{238}$U and $^{232}$Th, the procedure was developed using inductively coupled plasma-mass spectrometry (ICP-MS) with chemical extraction (see Ref. \cite{UTh} for more details). 

As shown in Table \ref{tab:level}, SK-Gd requires a method to determine low concentration $^{226}$Ra. 
However, the sensitivity for $^{226}$Ra of the previous method with the molecular recognition resin ``AnaLig-Ra01" was on the 1~mBq level \cite{Ra}. 
In addition, the recovery rate of $^{226}$Ra with the previous method was 81$\pm$4\% and the process time of the sample was 1~L per hour.  
To achieve the required sensitivity of SK-Gd, it was necessary to increase the concentration rate with a higher recovery rate and a shorter processing time. 
In this study, the procedure of chemical extraction was improved to achieve the sensitivity of on the  10$^{-1}$~mBq~kg$^{-1}$ level by solving these problems: the improved method was applied to SK-Gd to determine the concentration of $^{226}$Ra in Gd$_2$(SO$_4$)$_3{\cdot}$8H$_2$O.

\begin{table}[htbp]
\begin{center}
\caption{Summary of the maximum allowed level for SK-Gd and the typical values of commercially available product \cite{Pablo, icrc}. All units are mBq~kg$^{-1}$ (Gd$_2$(SO$_4$)$_3\cdot$8H$_2$O).}
\tabcolsep7pt\begin{tabular}{c|ccc} \hline
& $^{226}$Ra & $^{238}$U & $^{232}$Th \\ \hline
Requirement for SK-Gd & 0.5 & 5 & 0.05 \\
The typical concentration of  & \multirow{2}{*}{5} & \multirow{2}{*}{50} & \multirow{2}{*}{100} \\ 
commercially available product & & & \\ \hline
\end{tabular}
\label{tab:level}
\end{center}
\end{table}

\section{Experimental equipment}

\subsection{Chemical equipment}

\begin{figure}[htbp]
\centerline{\includegraphics[width=13cm]{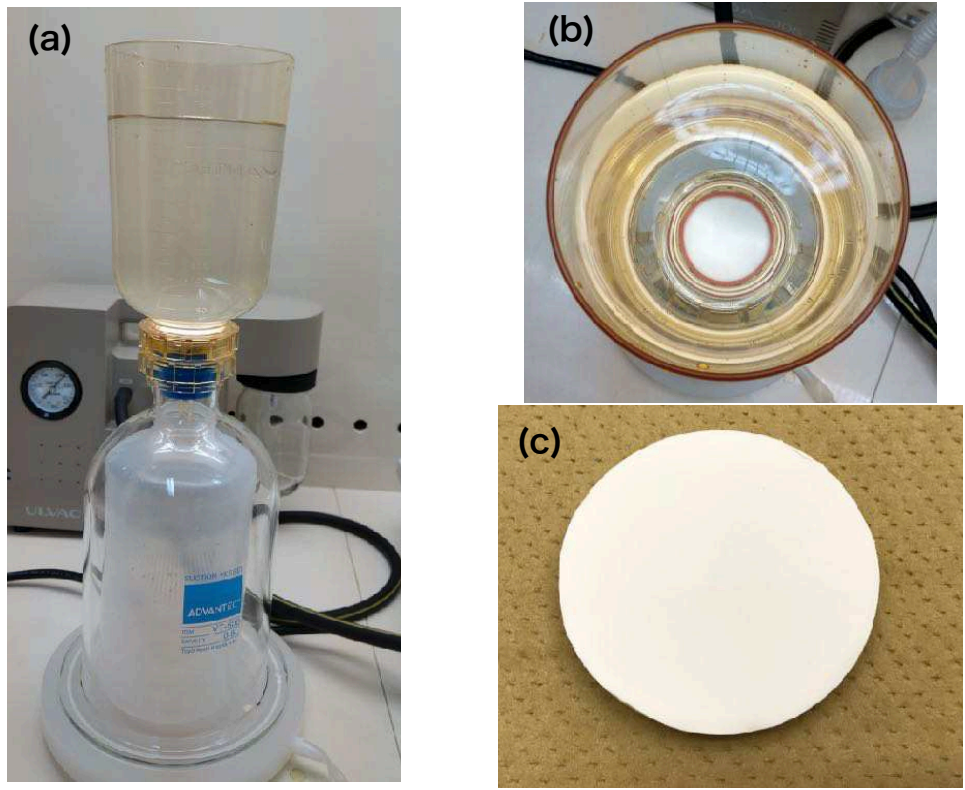}}
\caption{(A) Photograph of the entire experimental setup. (B) Top view of the setup. (C) Top view of the disk (47~mm diameter).}
\label{fig:SetUp}
\end{figure}

``Empore Radium Rad Disk" \cite{3M} was used for chemical extraction. 
The resin with the same chemical features as AnaLig-Ra01 was positioned on a filter (47~mm diameter and 0.5 $\mu$m thickness) made of polytetrafluoroethylene fibrils\footnote{This is generally called ``disk" and simply called disk in this paper from now on.}. 
Figure \ref{fig:SetUp} shows the experimental setup of the chemical extraction using a vacuum filtration system with the disk. 
The disk was placed on a holder with the volume of 800 mL (Advantech Toyo Ltd. \cite{Advantec}, KP-47), and the holder was connected to the vacuum container (Advantech Toyo Ltd. \cite{Advantec}, VT-500). 
The solution passes through the disk, and $^{226}$Ra in the solution is adsorbed by the resin bedded to the disk. 
The concentration of $^{226}$Ra can be determined by measuring the disk directly using an HPGe detector. 

To produce solutions with low contamination, ultra-pure SK water \cite{NIMA} was used. 
Electronic grade 70\% nitric acid (HNO$_3$) (Wako Pure Chemical Industries Ltd. \cite{Wako}) was used to wash the disk and efficiently dissolve Gd$_2$(SO$_4$)$_3{\cdot}$8H$_2$O in the SK water. 

To easily check the concentration of $^{226}$Ra in Gd$_2$(SO$_4$)$_3{\cdot}$8H$_2$O, $^{226}$Ra-rich hot spring water from the Kawakita hot spring in Ishikawa, Japan \cite{Tomita} was used as the calibration solution. 
The concentration of $^{226}$Ra in the hot spring water was 112$^{+34}_{-12}$ mBq L$^{-1}$, which was  determined by the HPGe detector measurement. 
The uncertainty was mainly due to the systematic uncertainty of the HPGe detector (+30\% or -10\%) and statistics \cite{Ra}. 
The sampled hot spring water was filtrated by membrane filters with the pore size of 0.45 $\mu$m and acidified to pH$\simeq$1 by HNO$_3$ for preservation. 

Because barium (Ba) and Ra have similar chemical features and ionic radii, Ba is frequently used as a tracer for Ra analysis to estimate the recovery rate \cite{RaBa}. 
Thus, 1000~mg~L$^{-1}$ Ba of standard solution (Merck Ltd. \cite{Merck}) was used to estimate the recovery rate. 
The details of studies for the recovery rate are described in Sec. \ref{sec:recovery}. 
The ICP-MS ``Agilent 7900" \cite{Agilent} was used to measure the concentration of Ba to estimate the recovery rate of $^{226}$Ra. 
The performance of this ICP-MS is described in Ref. \cite{UTh}. 

\subsection{High purity germanium detector and its detection efficiency}

\begin{figure}[htbp]
\centerline{\includegraphics[width=9cm]{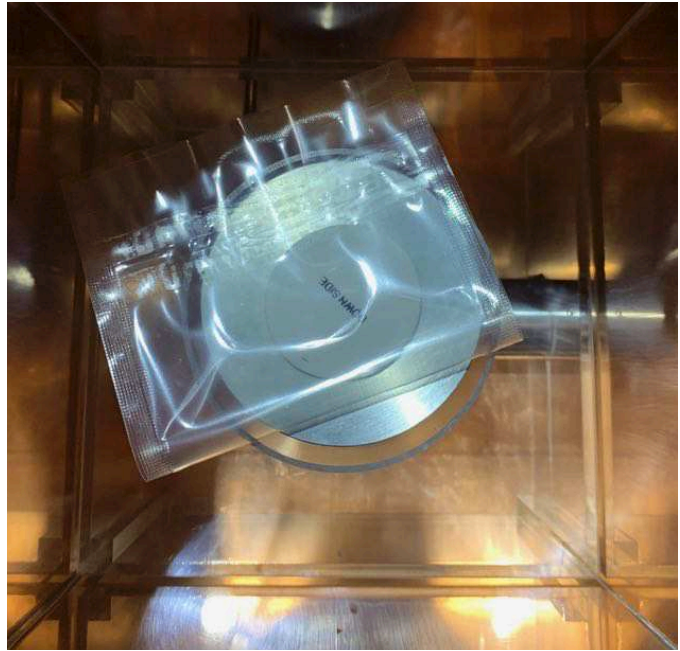}}
\caption{Setup for the disk measurement with the HPGe detector.}
\label{fig:GeDisk}
\end{figure}

The HPGe detector used for this measurement was a coaxial p-type HPGe crystal manufactured by CANBERRA France \cite{canberra}. 
The dimension of sample chamber was 23 $\times$ 23 $\times$ 48~cm$^3$. 
The details of the performance of the HPGe detector are described in Ref. \cite{Ra}.  
The samples measured by the HPGe detector were covered by an ethylene vinyl alcohol bag to keep radon  from samples inside the bag (Fig. \ref{fig:GeDisk}). 

\begin{figure}[htbp]
\centerline{\includegraphics[width=11cm]{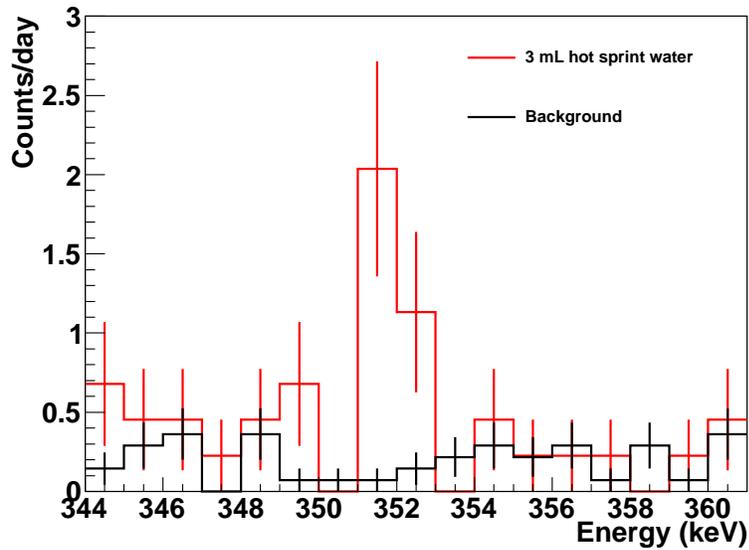}}
\caption{Energy spectra obtained by the HPGe detector around $^{214}$Bi 352 keV. The black line shows the background and red line indicates the extracted $^{226}$Ra from 3 mL of hot spring water (corresponding to 0.33~mBq). The error bar at each bin represents only statistical uncertainty.}
\label{fig:352keV}
\end{figure}

The concentration of $^{226}$Ra was evaluated using the characteristic $\gamma$-lines of $^{214}$Pb (609~keV) and $^{214}$Bi (352~keV and 1764~keV), which are daughter nuclei of $^{226}$Ra, by considering of their branching ratios and detection efficiencies. 
Figure \ref{fig:352keV} shows the typical observed energy spectra for $^{214}$Bi 352~keV measurements. 
The detection efficiency was evaluated by the Monte Carlo simulation \cite{geant4}. 
For example, the detection efficiency of 352~keV gamma rays originating from $^{214}$Bi, with the chemical extraction procedure using the disk was found to be 15.9\%. 
On the other hand, the detection efficiency of 352 keV gamma ray for $^{226}$Ra was determined to be 0.8\% for the direct measurement of 5~kg of  Gd$_2$(SO$_4$)$_3{\cdot}$8H$_2$O without a chemical extraction procedure.  
The detection efficiency was improved by a factor of 20  owing to the smaller volume of the sample using the disk. 

\section{Chemical extraction of $^{226}$Ra from Gd$_2$(SO$_4$)$_3{\cdot}$8H$_2$O and its performance \label{sec:recovery}}

The setup shown in Fig. \ref{fig:SetUp} was connected to a vacuum pump: the solution loaded into the holder could pass through the disk. 
The disk was initially washed by loading 50~mL of 3~mol~L$^{-1}$ HNO$_3$ and 50 mL of the ultra-pure SK  water into the holder. 
A total of 500~g Gd$_2$(SO$_4$)$_3{\cdot}$8H$_2$O was dissolved in 10~L of a 0.2~mol~L$^{-1}$ HNO$_3$ solution. 
Then, the sample solution was loaded into the holder, and the vacuum pressure was adjusted to produce a  flow rate of 50 mL min$^{-1}$ (3 L per hour, which is a three times faster processing time than that used in  the previous method \cite{Ra}). 
Then, the disk was directly placed on the HPGe detector and measured. 

The blank amount of $^{226}$Ra in the disk was evaluated by measuring 17 disks using the HPGe detector, the value of 1.9$^{+0.6}_{-0.4}$~mBq was obtained for 17 disks (corresponding to 0.11$^{+0.04}_{-0.02}$~mBq for each disk). 
In the previous method \cite{Ra}, the value of procedure blank was 0.3$\pm$0.2~mBq. 
This improved the detection limit.

The recovery rate of this procedure was evaluated using the $^{226}$Ra-rich hot spring water. 
The volume of 3 mL and 100 mL of the hot spring water was added to 10 L of a  Gd$_2$(SO$_4$)$_3{\cdot}$8H$_2$O dissolved sample solution. 
Then, the concentration of $^{226}$Ra adsorbed by the disk was measured using the HPGe detector for 4.5 days. 
As shown in Table \ref{tab:recovery}, the results of measurements were consistent with the expected amount of $^{226}$Ra, and the achieved sensitivity was on the $10^{-1}$~mBq level owing to the development of this  chemical extraction procedure. 

To estimate the recovery rate more accurately, Ba standard solution was added to the sample solution. 
The sample solution with the concentration of 4.0${\times}10^{-8}$ g (Ba) mL$^{-1}$ was loaded into the holder, and the solution (which passed through the disk) was collected and measured using the ICP-MS. 
The concentration of remaining Ba in the solution was $<0.1$\%. 
Those studies indicated that high matrix elements (Gd$_2$(SO$_4$)$_3{\cdot}$8H$_2$O) did not interfere with the extraction of $^{226}$Ra and the recovery rate obtained using the developed method was estimated to be $>$99.9\%. 
This is a significant improvement from the recovery rate obtained using the previous method (81$\pm$4\%) \cite{Ra}."

Table \ref{tab:performance} shows the comparison of performance between the previous method in Ref.  \cite{Ra} and the developed method in this study. 
This result shows the sensitivity of the improved method is on the 10$^{-1}$~mBq level and sufficient for  measuring the experimentally allowed level of $^{226}$Ra for SK-Gd. 

\begin{table}[htbp]
\begin{center}
\caption{Summary of the study for the recovery rate. The blank of the disk was already subtracted.}
\tabcolsep7pt\begin{tabular}{c|cc} \hline
Hot spring water & Expected amount & Results \\
(mL) & of $^{226}$Ra (mBq) & (mBq) \\ \hline
3 & 0.33$^{+0.10}_{-0.04}$ & 0.4${\pm}0.2$  \\
100 & 11.2$^{+3.4}_{-1.2}$ & 11.3$^{+3.4}_{-1.1}$  \\ \hline
\end{tabular}
\label{tab:recovery}
\end{center}
\end{table}

\begin{table}[htbp]
\begin{center}
\caption{Summary of the performance in the previous and developed methods.}
\tabcolsep7pt\begin{tabular}{c|cc} \hline
 & Previous method & Developed method \\ \hline
Recovery rate (\%) & 81$\pm$4 & $>$99.9 \\
Process time for sample (L per hour) & 1  & 3 \\
Procedure blank (mBq) & 0.3$\pm$0.2 & 0.11$^{+0.04}_{-0.02}$ \\
Sensitivity of the test sample (mBq) & 0.9$\pm$0.5 & 0.4$\pm$0.2 \\ \hline
\end{tabular}
\label{tab:performance}
\end{center}
\end{table}

\section{Application to SK-Gd and results of measuring Gd$_2$(SO$_4$)$_3{\cdot}$8H$_2$O}

SK-Gd will be conducted in several experimental phases. 
For the first experimental phase, 13 tons of Gd$_2$(SO$_4$)$_3{\cdot}$8H$_2$O will be dissolved in the SK tank corresponding to a 50\% neutron tagging efficiency \cite{EGADS}. 
Gd$_2$(SO$_4$)$_3{\cdot}$8H$_2$O was produced with many lots: thus, all the lots should be measured before loading. 
Currently, 14 tons of Gd$_2$(SO$_4$)$_3{\cdot}$8H$_2$O are being measured using the developed method  to confirm that their radio impurities are below the experimentally allowed levels (see Table \ref{tab:level}). 
The concentration of $^{226}$Ra in Gd$_2$(SO$_4$)$_3{\cdot}$8H$_2$O (unit: mBq~kg$^{-1}$) can be obtained from the amount of $^{226}$Ra measured in the disk divided by the weight of Gd$_2$(SO$_4$)$_3{\cdot}$8H$_2$O. 
Table \ref{tab:RaGd} shows the results of the measurement of $^{226}$Ra in Gd$_2$(SO$_4$)$_3{\cdot}$8H$_2$O for several production lots determined using the improved chemical extraction method. 
The signals of $^{214}$Pb and $^{214}$Bi were not observed above the statistical uncertainty. 
The concentrations of $^{226}$Ra in the measured products were confirmed to be below the experimentally allowed level. 

On the basis of these studies and measurements, a highly sensitive method for measuring low concentration  $^{226}$Ra was established. 
This method can be applied to other non-accelerator particle physics experiments as well. 
For example, the XENON-1T detector will be upgraded to the XENON-nT detector with a neutron veto system  which is based on a high-purity Gd-loaded water Cherenkov detector \cite{Xenon}.

\begin{table}[htbp]
\begin{center}
\caption{Summary of the measurements of Gd$_2$(SO$_4$)$_3{\cdot}$8H$_2$O. The upper limits represent 90\% confidence level.}
\tabcolsep7pt\begin{tabular}{ccc} \hline
Lot No. & Concentration of $^{226}$Ra & Measurement time \\
 & (mBq kg$^{-1}$) & (day) \\ \hline
1 & $<0.4$ & 6.0\\
2 & $<0.3$ & 11.0\\
3 & $<0.3$ & 8.8\\ 
4 & $<0.2$ & 9.6\\ 
5 & $<0.5$ & 8.7\\ 
6 & $<0.2$ & 13.0\\ \hline
\end{tabular}
\label{tab:RaGd}
\end{center}
\end{table}

\section{Conclusion}

The method for measuring $^{226}$Ra using an HPGe detector with chemical extraction was improved to determine low concentration $^{226}$Ra in a high matrix sample. 
More than 99.9\% of $^{226}$Ra was recovered from the high matrix sample with a shorter processing time of the chemical extraction, which resulted in the sensitivity on the 10$^{-1}$~mBq level. 
The improved method is being applied to SK-Gd to determine $^{226}$Ra in Gd$_2$(SO$_4$)$_3{\cdot}$8H$_2$O. 
Currently, all the measured Gd$_2$(SO$_4$)$_3{\cdot}$8H$_2$O products, which will actually be loaded into the SK tank, were confirmed to be below than the maximum allowed $^{226}$Ra concentration level.




\begin{thebibliography}{99}

\bibitem{Sr}
S. Dulansk$\acute{\rm a}$, V. Gardonov$\acute{\rm a}$, B. Remenec, L. M$\acute{\rm a}$tel, and H. Cab$\acute{\rm a}$nekov$\acute{\rm a}$, J. Radioanal. Nucl. Chem. 309, 853 (2016).

\bibitem{IBC}
IBC Advanced Technologies, Inc., AnaLig Products (IBC, American Fork, UT, 2018) (available at:
http://www.ibcmrt.com/products/analig/).

\bibitem{3M}
Technical information of Empore Radium Rad Disk available at http://www.envexp.com/pdf/Empore\%20Rad\%20Disks.pdf.

\bibitem{Disk1}
A. $\check{\rm D}$urecov$\acute{\rm a}$ et al., Czech. J. Phys. 49/S1 (1999).

\bibitem{Disk2}
M. Cook and R. Kleinschmidt, Aust. J. Chem. 2011, 64, 880–884. 

\bibitem{Disk3}
J. Fons-Castells et {\it al}., Applied Radiation and Isotopes 124, 2017, 83-89.

\bibitem{NIMA}
S. Fukuda et al., Nucl. Instrum. Methods. A501, (2003), 418.

\bibitem{Gd}
John F. Beacom and Mark R. Vagins, Phys. Rev. Lett. 93, 171101 (2004).

\bibitem{EGADS}
Ll. Marti {\it et al.}, Nucl. Instr. Meth. A, {\bf 959},  (2020) 163549.

\bibitem{SRN}
S. Horiuchi et al., Phys. Rev. D79, 083013 (2009).

\bibitem{solar}
K. Abe et al., Phys. Rev. D, 94, 052010, (2016).

\bibitem{Pablo} 
P. Ferna, PhD Thesis, University Autonomous of Madrid, Mar. 2017 (available at http://www-sk.icrr.u-tokyo.ac.jp/sk/\_pdf/articles/2017/thesis\_PabloFernandez\_March2017.pdf)

\bibitem{icrc}
Shintaro Ito, PoS ICRC2017 (available at https://pos.sissa.it/301/1012/pdf).

\bibitem{UTh}
S. Ito et al., Prog. Theor. Exp. Phys. 113H01 (2017). 

\bibitem{AnaLig}
IBC (available at http://www.ibcmrt.com/products/analig/)

\bibitem{Ra}
S. Ito et al., Prog. Theor. Exp. Phys. 091H01 (2018). 

\bibitem{Advantec}
Advantec Toyo Ltd. (available at http://www2.advantec.co.jp/en/).

\bibitem{Wako}
Wako Pure Chemical Ltd. (available at http://www.wako-chem.co.jp/)

\bibitem{Tomita}
J. Tomita et al., J. Hot Spring Sci., 58, 241-255 (2009). 

\bibitem{RaBa}
International Atomic Energy Authority, Analytical Methodology for the Determination of Radium Isotopes in Environmental Samples. IAEA/AQ/19. (IAEA, Vienna, 2010) (available at: https://www-pub.iaea.org/MTCD/Publications/PDF/IAEA-AQ-19\_web.pdf).

\bibitem{Merck}
Merck Ltd. (available at http://www.merck.co.jp/en/company/merck\_ltd/merck\_ltd.html)

\bibitem{Agilent}
Agilent Technologies (available at https://www.agilent.com/en-us/products/icp-ms/icp-ms-systems/7900-icp-ms)

\bibitem{canberra}
CANBERRA France Inc. (available at http://www.canberra.com/).

\bibitem{geant4}
S. Agostinelli, {\textit{et al}.}, Nucl. Instr. Meth. A, {\bf 506},  (2003) 250-303.

\bibitem{Xenon}
The XENON Experiment (available at: https://science.purdue.edu/xenon1t/?p=1051)

\end{thebibliography}
%


\ack
This work was supported by the JSPS KAKENHI Grants Grant-in-Aid for Scientific Research on Innovative Areas No. 26104004, 26104006, 19H05807, and 19H05808, Grant-in-Aid for Specially Promoted Research No. 26000003, Grant-in-Aid for Young Scientists No. 17K14290, and Grant-in-Aid for JSPS Research Fellow No. 18J00049.


\end{document}